# Chemically stabilized epitaxial wurtzite-BN thin film

Badri Vishal,[1] Rajendra Singh,[1] Abhishek Chaturvedi,[2] Ankit Sharma,[1] M.B. Sreedhara,[1] Rajib Sahu,[1] Usha Bhat,[1] Upadrasta Ramamurty,[2] and Ranjan Datta[1,*]

[1]*International Centre for Materials Science, Chemistry and Physics of Materials Unit, Jawaharlal Nehru Centre for Advanced Scientific Research, Bangalore 560064, India.*

[2]*Department of Materials Engineering, Indian Institute of Science, Bangalore 560012, India.*

We report on the chemically stabilized epitaxial *w*-BN thin film grown on *c*-plane sapphire by pulsed laser deposition under slow kinetic condition. Traces of no other allotropes such as cubic (*c*) or hexagonal (*h*) BN phases are present. Sapphire substrate plays a significant role in stabilizing the metastable *w*-BN from *h*-BN target under unusual PLD growth condition involving low temperature and pressure and is explained based on density functional theory calculation. The hardness and the elastic modulus of the *w*-BN film are 37 & 339 GPa, respectively measured by indentation along <0001> direction. The results are extremely promising in advancing the microelectronic and mechanical tooling industry.

*Corresponding author e-mail: ranjan@jncasr.ac.in

Boron Nitride (BN) exhibits numerous allotropes e.g., hexagonal (*h*-BN, P6$_3$/mmc), rhombohedral (*r*-BN, R3m), cubic (*c*-BN, Fd3m) and wurtzite (*w*-BN, P6$_3$mc) which are analogous to Carbon (C) allotropes. Among these, *h*-BN is the most stable form found at room temperature and pressure. *c*-BN and metastable *w*-BN can be stabilized, but require extreme temperature (1730-3230 °C) and pressure (5-18 GPa) condition.[1] The equilibrium phase diagram of BN depicts that the stabilization of *w*-BN requires either *c*-BN or *h*-BN as the starting phase and may be difficult to synthesize in pure form.[2] Observation of synthetic *w*-BN, converted from *h*-BN at pressure of 11.5 GPa and temperature of 2000K, was first reported in 1963 by Bundy and Wentorf Jr.[3] Subsequently, Various methods e.g., static high-pressure, shock-wave compression method, direct conversion from *h*-BN have been reported.[4-8] Experimental investigation into the properties of *w*-BN are scarce because of the difficulty in synthesizing sufficiently large and pure crystals of it. Recently, synthesis of 2 mm diameter and 350 µm thick *w*-BN crystals with 98% purity (*h*-BN is the residue phase) under direct conversion method involving high temperature (1500 °C) and pressure (4 GPa) was reported.[7] There are reports on the formation of *w*-BN and *c*-BN by thin film growth procedure.[9-13] Polycrystalline (2-20 µm grain size) *w*-BN film is deposited on amorphous C film by pulsed laser deposition.[11] Only *c*-BN thin film is deposited on Si (100) substrate by ion assisted pulse laser deposition at 400 °C and 10$^{-5}$ Torr.[12] Mixture of *c*-BN and *w*-BN has also been reported by PLD on WS$_2$ and ReS$_2$ template.[14]

Two-dimensional (2D) layered *h*-BN (also called white graphene) has recently attracted considerable attention to improve the performance of graphene, other 2D materials such as MoS$_2$ and also being explored as an active material for the optoelectronic and energy applications.[15-16] On the other hand, *c*-BN and *w*-BN are attractive due to high hardness and the potential in electronic applications remains unexplored.[14,17] In the *w*-BN structure each atom is tetrahedrally coordinated with a B-N bond length of 157 pm and a bond angle of 109.5° with '..*aa'bb'aa'*..' stacking of the basal plane along *c* direction and the structure is denser and harder compared to *h*-BN.[18] Theoretical predictions indicate that both *c*-BN and *w*-BN are super hard, with *w*-BN being the second most hard material (114 GPa) after Lonsdaleite (hexagonal C, hardness - 152 GPa), due an intermediate bond flipping structural phase transition.[19] Recently, another new form of C called Q-carbon has been predicted to show higher hardness than diamond.[20] However, the unavailability of sufficiently large crystal of *w*-BN and Lonsdaleite prevented experimental verification of these predictions.[19,21,22] Therefore, diamond retained the top position in the list. Both *w*-BN and *c*-BN have the advantage over diamond in terms of excellent chemical and thermal stability (oxidation temperature is 1300 °C compared to 700 °C of diamond) suitable for high speed manufacturing. Experimentally, the nanocrystalline (~14 nm) composite of *c*-BN and *h*-BN exhibited hardness of 85 GPa which is close to that of diamond.[23] The millimeter size bulk *w*-BN crystal (98% purity) showed Hardness and Young modulus values of ~ 54 and 860 GPa, respectively.[7]

Here, we report on the formation of large area single phase epitaxial thin film of *w*-BN at much lower temperature and pressure compared to the extreme condition usually require for stabilizing this metastable phase.[2] The film is grown by pulsed laser deposition (PLD) under slow kinetic condition. The formation of *w*-BN phase is confirmed by high resolution transmission electron microscopy (HRTEM), X-ray diffraction (XRD) and Raman spectroscopy. The stabilization of metastable *w*-BN at low temperature (400 °C) and pressure ($10^{-5}$ Torr) combination is rationalized based on chemical interaction between the O-planes of *c*-plane sapphire and the B atoms of the first *h*-BN layer, which is supported by first principle calculation. This chemical interaction initiates staggering in the first *h*-BN layer which subsequently transforms into *w*-BN.

Nanoindentation on the thin films show that $H$ and $E$ of this phase are 37 and 339 GPa, respectively.

Thin films of $w$-BN are grown by pulsed laser deposition. on 8×8 mm$^2$ size 'c' plane sapphire substrate. The target pellet is prepared from $h$-BN powder obtained from Sigma Aldrich (99.9%) by first cold pressing and then sintering at 800 °C for 5 h in a vacuum chamber (~10$^{-5}$ Torr). The ablation frequency used is 1 Hz during the film growth. The slow laser ablation rate allows sufficient time for kinetic relaxation of the nucleation layer to establish epitaxial relationship with the underlying substrate. This also helps in eliminating misaligned crystallites for highly lattice mismatched epitaxy.[24, 25] The pressure is kept constant at ~10$^{-5}$ Torr throughout the growth schedule while the temperature of the growth was 400 and 800 °C for the nucleation and final growth, respectively. The formation of epitaxial $w$-BN thin film is confirmed by XRD, HRTEM and electron diffraction techniques. TEM cross sectional samples are prepared by first mechanical polishing and then Ar ion milling to perforation around which large electron transparent thin area is generated. Raman spectra are recorded using a custom-built Raman spectrometer using a 532 nm laser excitation and a grating of 1800 lines/mm at room temperature.[26] The laser power at the sample is approximately 1 mW.

To measure the mechanical properties of the thin film, nanoindentation is performed on 100 nm thick BN film on sapphire substrate using Hysitron Triboindenter, Minneapolis, MN, USA, equipped with a three-sided pyramidal Berkovich diamond indenter with a tip radius of ~100 nm. The equipment records the load, P, and depth of penetration, h, of the indenter with resolutions of 1 nN and 0.2 nm, respectively. The tests are performed both under displacement and load controlled modes to examine the reproducibility of the results since the maximum depths of penetrations, $h_{max}$, are small. Around 20 indentations are performed on each sample. The recorded P-h responses are analyzed using the standard Oliver-Pharr method to extract the reduced elastic modulus, $E_r$, and hardness, H, of the sample.[27]

To understand the substrate effect on the phase transition from *h*-BN to *w*-BN theoretical calculations are performed based on density functional theory (DFT) using self-consistent plane wave pseudo potential as implemented in Quantum Espresso (QE) code.[28] The ionic core-valence electron interactions are modeled using ultra-soft pseudo-potentials.[29] Electronic exchange-correlation energy is approximated using Perdew-Burke-Ernzerhof (PBE) functional within generalized gradient approximation (GGA).[30] The following cases were simulated, (*i*) monolayer *h*-BN (*ii*) Modulated monolayer BN, (*iii*) bilayer *h*-BN and (*iv*) modulated bilayer BN on top of O-end $Al_2O_3$ with epitaxial relation *w*-BN $<11\bar{2}0> \parallel Al_2O_3 <01\bar{1}0>$ that were observed experimentally. The schematic of simulated structure is shown in Fig.1 and supplementary document [Fig.S1]. Kinetic energy and augmented charge density cutoffs used are 50 and 400 Ry, respectively. Atomic positions and cell parameters are fully relaxed below energy convergence of $10^{-5}$ eV. Optimized structures of BN and $Al_2O_3$ are stacked according to epitaxial relation and then VC(variable cell)-relaxation is carried out with 8×8×8 k-mesh according to the scheme proposed by Monkhorst and Pack.[31] Relaxed supercell parameters of the *h*-BN on $Al_2O_3$ substrate is $a = b = 4.840$ Å and $\alpha = \beta = 90°$, $\gamma = 120°$ and additional 20 Å of vacuum is created along *c*-direction on top of BN.

We begin with the description of structural characterization confirming the formation of epitaxial *w*-BN thin film on (0001) plane of sapphire. From the low magnification TEM image uniform and smooth film of *w*-BN with thickness ~ 20 nm can be observed [Fig. 2(a)]. The film can be grown as thick as required following the slow kinetic condition. Indentation experiment was carried out on a film of 100 nm thickness. Fig. 2(b) & (c) are the electron diffraction (ED) pattern and schematic of in-plane orientation relationship between *w*-BN and sapphire, respectively. The epitaxial relationship is found to be *w*-BN [01$\bar{1}$0] ∥ α-Al$_2$O$_3$ [11$\bar{2}$0] and the relative rotation of *c* axis between film and substrate is 30°, which is similar to the growth of GaN or ZnO on *c*-plane sapphire. The chemical interaction between B and O atoms is responsible for such an arrangement and is explained latter with the aid of DFT.

The Raman spectra of the *w*-BN thin film is given in Fig. 3. The Raman spectra of reference powder *h*-BN, *c*-BN and *w*-BN are provided in the supplementary document [Fig. S2].[23] The peak, observed around 1370 cm$^{-1}$ corresponds to the *h*-BN structure. For *c*-BN, two distinct sharp peaks 1057 and 1309 cm$^{-1}$ were noted, whereas many broad peaks are present in case of *w*-BN. The FWHM of the Raman peaks corresponding to *w*-BN are relatively narrower in the present study, as compared to those reported earlier, which is due to large crystalline area in the former. During the study, it is also found that BN grows with different polytypic phases on TMDs depending on the TMD template (MoS$_2$, WS$_2$, and ReS$_2$).[14] XRD and XPS spectra supporting the formation of *w*-BN can be found in the supplementary document [Fig. S3].

HRTEM images of *w*-BN thin film on *c*-plane sapphire (Al$_2$O$_3$) along two different zone axes (Z.A.) orientations are given in Fig. 4. The film is relaxed and the lattice parameters of *w*-BN are approximately *a* = 2.58 Å and *c* = 4.29 Å. The atomic registry of the film with the terminating O-plane of *c*-plane sapphire substrate and specific stacking of atomic planes '…*ababa*..' can clearly be observed and schematics are provided for guidance. In addition to the ED pattern, the specific stacking of atomic planes along the growth direction in the film also confirming the formation of *w*-BN phase. Presence of *h*-BN or any other allotropes is not detected and *w*-BN is the only phase formed right from the film-substrate interface with some planar faults present in the film. This suggests that the phase transformation from $h$-BN to $w$-BN must have occurred at the O-terminated (0001) plane of sapphire via chemical interaction which is further supported by DFT based calculation.

The difference in electronegativity between B and O is high compared to N and O atoms. Therefore, the chemical bonding between BN to sapphire is due to bonding between B and O atoms which is energetically more favourable compared to N and O bonding by 173 meV/atom. This is indicated with the colour round circles in Fig. 4. From the HRTEM image, the inter atomic distance between B and O atoms along projected $c$ direction is ~1.34 Å. The shortest inter-atomic distances along $c$ direction for $w$-BN and $h$-BN are ~1.4 and 2.6 Å, respectively. This suggests that the bonding between the first BN layer to O-plane of sapphire is due to covalent chemical bonding and not weak van der Waals interaction. Moreover, the structure of $h$-BN can be converted to $w$-BN by staggering the layer by selective bonding between B and O atoms as shown in Fig. 1(a). HRTEM image also shows that the first layer is like staggered $h$-BN [Fig. 4]. As already mentioned, $h$-BN is used as starting compound during the film growth. Theoretically similar structural transformation through staggering the $h$-BN to $w$-BN was considered under pure shear stress and not by chemical interaction.[21]

In view of above, DFT calculation were performed for both monolayer, bi-layer $h$-BN and staggered $h$-BN on O-terminated sapphire to get insights into the stability and chemical interaction through charge transfer between B and O atoms leading to structural phase transition from $h$-BN to $w$-BN. The structure of sapphire (α-$Al_2O_3$, space group $R\bar{3}c$) consists of 12 Al and 18 O atoms per unit cell with lattice parameter $a = 4.758$ Å, $c = 12.992$ Å. In O-ending (0001) plane of $Al_2O_3$, there are two distinct O-O-O equilateral triangles with O-O interatomic distances 2.52 and 2.86 Å where B-B-B tringle of staggered *h*-BN or the first layer of (0001) *w*-BN stack with a B-B interatomic distance 2.58 Å. This gives +2.3% and -9.7 % strain in the BN lattice corresponding to two different O-O-O tringles and an overall -6.1 % lattice mismatch along *w*-BN $< 01\bar{1}0 > \parallel$ Al$_2$O$_3$ $< 2\bar{1}\bar{1}0 >$ resulting in non-uniform bond length [Fig. 2(c)]. The calculation revealed that *w*-BN on sapphire with B-O bonding is more stable (7.76 eV/atom) than *h*-BN (7.55 eV/atom) for the monolayer coverage. Further calculation shows that the staggering propagates to the bi-layer as a stable system (7.92 eV/atom) compared to bi-layer flat *h*-BN (7.83 eV/atom) and eventually form the wurtzite structure [Fig. 1(b)]. The valence charge density plot reveals charge transfer between O and B atoms for staggered *h*-BN [Fig. 1 (c) & (d)] but not for the flat *h*-BN [Fig. S1]. The observation by HRTEM imaging at the interface and combination of first principle calculation suggests that the selective and stronger chemical interaction between the O atoms and B atoms in *h*-BN for the first deposited layer initiated the structural phase transition from *h*-BN to *w*-BN and subsequently propagated along the (0001) direction.

A representative *P-h* (load vs. penetration depth) response obtained through nano-indentation on the *w*-BN thin film is displayed in Fig. 5. A prominent discrete displacement jumps or 'pop-in' at $P \approx 600$ μN, which corresponds a *h* of ~20 nm can be seen. Such pop-ins can be either due to cracking or delamination of the film, both of which make the film compliant. Therefore, we confined subsequent experiments to extremely shallow depths ($h_{max}$ = 15 nm). Representative *P-h* response obtained in those experiments is displayed as the inset of Fig. 5. No pop-ins can be seen. The E and H values obtained from such experiments are 339 and 37 GPa, respectively. In addition to calibration before each of set of nanoindentations, which are performed by fused quartz specimens, nano-indentation on bare α Al$_2$O$_3$ was also performed. Results of these experiments [Fig. S5] are in good agreement with those reported in literature [E = 444.4±20.1 GPa and H = 28.9±2.3 GPa along (0001)].[32] Importantly, the E values obtained via shallow depth nano-indentations ($h_{max}$ = 40 nm), are similar to those obtained with much higher $h_{max}$. This observation confirmed the validity of extracting E and H values from shallow depth indentations.

The experimental values of hardness (37 GPa) and elastic modulus (339 GPa) of the present *w*-BN thin film is in the lower side among various experimental measurement of this phase either in bulk or nano-composite form reported earlier [Table 1]. The present measurement of epitaxial *w*-BN thin film involves indentation along <0001> direction due to the orientation of film and such measurement is highly anisotropic compared to the earlier reports. Note that the <0001> direction is not the closed packed direction of *w*-BN structure and hence unlikely to exhibit the highest stiffness or hardness.[19] Theoretically, the tensile strength of *w*-BN is 68/90 GPa along <0001> and $< 11\bar{2}0 >$ directions, respectively.[19,22] The bi-axial shear strength with intermediate structural transition can reach 114 GPa. If we extrapolate the experimental data based on the theory calculation then the film would show strength of ~ 65 GPa along ($< 11\bar{2}0 >$) direction, which is still lower than predicted. Possible reasons for the experimental values being less that the theoretical predictions are that (a) the structure may not have undergone bond flipping intermediate structural transitions proposed earlier, and (b) the presence of planar defects in the as-grown film whereas theoretical predictions are made on perfect crystals and hence are ideal values. For diamond, the theory and experimental hardness closely matches may be due to diamond crystals are near perfect with less influence from the defects.

In conclusion, *w*-BN thin film is grown under relatively low pressure and temperature condition in PLD compared to the extreme higher temperature and pressure condition usually required to stabilize this metastable phase. The phase transition occurred due to chemical interaction between B and O atoms leading to staggering in flat *h*-BN layer transforming to *w*-BN structure. The hardness of the film is 37 GPa lower than theoretical prediction due to indention along soft direction, thin film geometry and defects present in the film. Nonetheless, the growth of such epitaxial thin film may find application in tooling and microelectronic industry.

**Supplementary Information**

Supplementary documents contain information on reference Raman spectra, additional HRTEM images, X-ray and XPS spectra, theoretical schematic and charge density plot of *h*-BN on sapphire.

**Acknowledgement**

The authors at JNCASR sincerely acknowledge ICMS for funding.

**References**

[1]R. H. Wentorf Jr, J. Chem. Phys. **34**, 809 (1961).

[2]V. L. Solozhenko, D. Häusermann, M. Mezouar, and M. Kunz, Appl. Phys. Lett. **72**, 1691 (1998).

[3]F.P. Bundy and R.H. Wentorf Jr, J. Chem. Phys. **38**, 1144 (1963).

[4]E. Tani, T. Sōma, A. Sawaoka, and S. Saito, Jpn. J. Appl. Phys. **14**, 1605 (1975).

[5]T. Sōma, A. Sawaoka and S. Saito, Mat. Res. Bull. **9**, 755 (1974).

[6]T. Akashi, H. R. Pak, and A. B. Sawaoka, J. Mater. Sci. **21**, 4060 (1986).

[7]M. Deura, K. Kutsukake, Y. Ohno, I. Yonenaga and T. Taniguchi, Jpn. J. Appl. Phys. **56**, 030301 (2017).

[8]F. R. Corrigan and F. P. Bundy, J. Chem. Phys. **63**, 3812 (1975).

[9]M. Sokołowski, J. Cryst. Growth **46**, 136 (1979).

[10]J. Szmidt, A. Jakubowski, A. Michalski, and A. Rusek, Thin Solid Films **110**, 7 (1983).

[11]G. Kessler, H-D Bauer, W. Pompe and H-J Scheibe, Thin Solid Films **147**, L45 (1987).

[12]D. L. Medlin, T. A. Friedmann, P. B. Mirkarimi, P. Rez, M. J. Mills and K. F. McCarty, J. Appl. Phys. **76**, 295 (1994).

[13]T. K. Paul, P. Bhattacharya, and D. N. Bose, Appl. Phys. Lett. **56**, 2648 (1990).

[14]U. Bhat, R. Singh, B. Vishal, A. Sharma, H. Sharona, R. Sahu and R. Datta, preprint arXiv:1710.04160 (2017).

[15]G. Giovannetti, P. A. Khomyakov, G. Brocks, P. J. Kelly and Van Den Brink, Phys. Rev. B **76**, 073103 (2007)

[16]M. R. Gao, Y. F. Xu, J. Jiang and S. H. Yu, Chem. Soc. Rev. **42**, 2986 (2013).

[17]N. Izyumskaya, D. O. Demchenko, S. Das, Ü. Özgür, V. Avrutin, H. Morkoç, Adv. Electron. Mater. **3**, 1600485 (2017)

[18]R. R. Wills, Int. J. High Technology Ceramics **1**, 139 (1985).

[19]Z. Pan, H. Sun, Y. Zhang, and C. Chen, Phys. Rev. Lett. **102**, 055503 (2009).

[20]J. Narayan and A. Bhaumik, J. Appl. Phys. **118**, 215303 (2015).

[21]W. J. Yu and W. M. Lau, Phys. Rev. B **67**, 014108 (2003).

[22]R.F. Zhang, S. Veprek and A. S. Argon, Phys. Rev. B **77**, 172103 (2008).

[23]N. Dubrovinskaia, V. L. Solozhenko, N. Miyajima, V. Dmitriev, O. O. Kurakevych and L. Dubrovinsky, Appl. Phys. Lett. **90**, 101912 (2007).

[24]B. Loukya, P. Sowjanya, K. Dileep, R. Shipra, S. Kanuri, L. S.Panchakarla, and R. Datta, J. Cryst. Growth **329**, 20 (2011).

[25]R. Sahu, D. Radhakrishnan, B. Vishal, D. S. Negi, A. Sil, C. Narayana and R. Datta, J. Cryst. Growth **470**, 51 (2017).

[26]G. V. P. Kumar and C. Narayana, Curr. Sci. **93,** 778-781 (2007).

[27]W. C. Oliver and G. M. Pharr, J. Mater. Res. **7**, 1564 (1992).

[28]P. Giannozzi, S. Baroni, N. Bonini, M. Calandra, R. Car, C. Cavazzoni, D. Ceresoli, G. L. Chiarotti, M. Cococcioni, and I. Dabo, J. Phys.: Condens. Matter **21**, 395502 (2009).

[29]D. Vanderbilt, Phys. Rev. B **41**, 7892 (1990).

[30]J. P. Perdew, K. Burke, and M. Ernzerhof, Phys. Rev. Lett. **77**, 3865 (1996).

[31]H. J. Monkhorst and J. D. Pack, Phys. Rev. B **13**, 5188 (1976).

[32]S. Ruppi, A. Larsson, and A. Flink, Thin Solid Films **516**, 5959 (2008).

[33]F. Gao, J. He, E. Wu, S. Liu, D. Yu, D. Li, S. Zhang and Y. Tian, Phys. Rev. Lett. **91**, 015502 (2003).

[34]A. Šimůnek, J. Vackář, Phys. Rev. Lett. **96**, 085501 (2006).

[35]N. Dubrovinskaia, S. Dub and L. Dubrovinsky, Nano Lett. **6**, 824 (2006).

[36]R.A. Andrievski, Int. J. Refrat. Met. Hard Matt. **19**, 447 (2001).

[37]Y. Tian, B. Xu, D. Yu, Y. Ma, Y. Wang, Y. Jiang, W. H. C. Tang, Y. Gao, K. Luo, Z. Zhao, L.-M. Wang, B. Wen, J. He and Z. Liu, Nature **493**, 385 (2013).

[38]H. Sumiyaa and T. Irifuneb, Diam. Relat. Mater. **13**, 1771 (2004).

**Table I**. List of reported hardness (H) and modulus (E) values of known hard materials obtain from experiment (Exp.) or calculations (Calc.). Experimental measurement methods i.e. Vickers (V), Knoop (K) is indicated.

| Phase | Phase | H (GPa) | E (GPa) | Exp. Or calc. |
|---|---|---|---|---|
| *w*-BN | Thin Film | 37 (V) | 339 | Exp.(present work) |
| Diamond-C | Bulk | 96(V)[20,36] | 535[36] | Exp. |
| Natural Diamond | Bulk | 57-104 (K)[35] | 1140[35] | Exp. |
| Diamond | Nanocrystallite | 105 (K)[35], 120-145 (K)[38] | 1070[35] | Exp. |
| Lonsdelite | | 152[19] | | Calc. |
| c-BN | Bulk | 64 (V)[33] | | Calc. |
| c-BN | Nanotwins in bulk | 108-196 (V)[37] | | Exp. |

**Figures:**

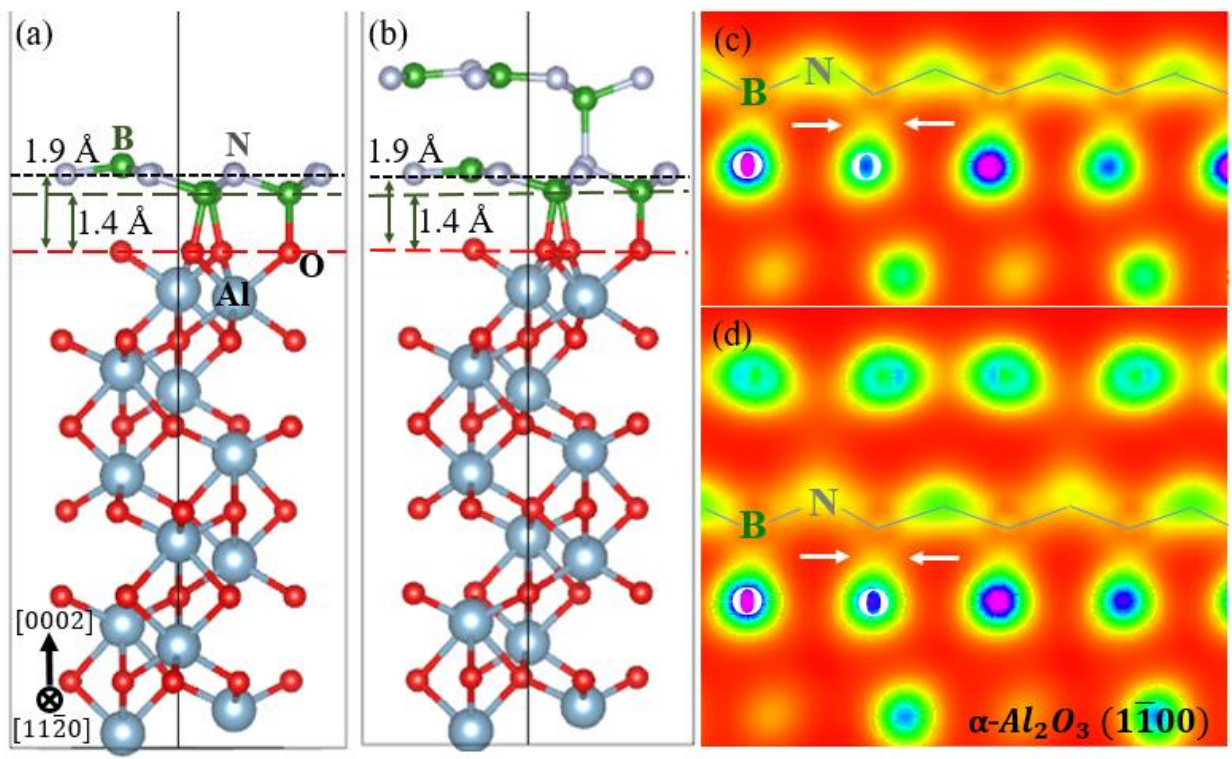

**Figure 1.** Schematic structure of monolayer (a) staggered *h*-BN on *c*-plane sapphire. The stability of staggered *h*-BN configuration is 210 meV/atom more compared to flat *h*-BN. (b) staggered bi-layer *h*-BN showing nucleation and growth of *w*-BN phase along *c* direction. Charge density plot showing additional interaction between B and O atoms at the interface in both (c) mono and (d) bi-layer leading to staggering configuration of flat *h*-BN. (Schematic structure and Charge density plot of flat mono and bi-layer *h*-BN is shown in supplementary Fig. S1).

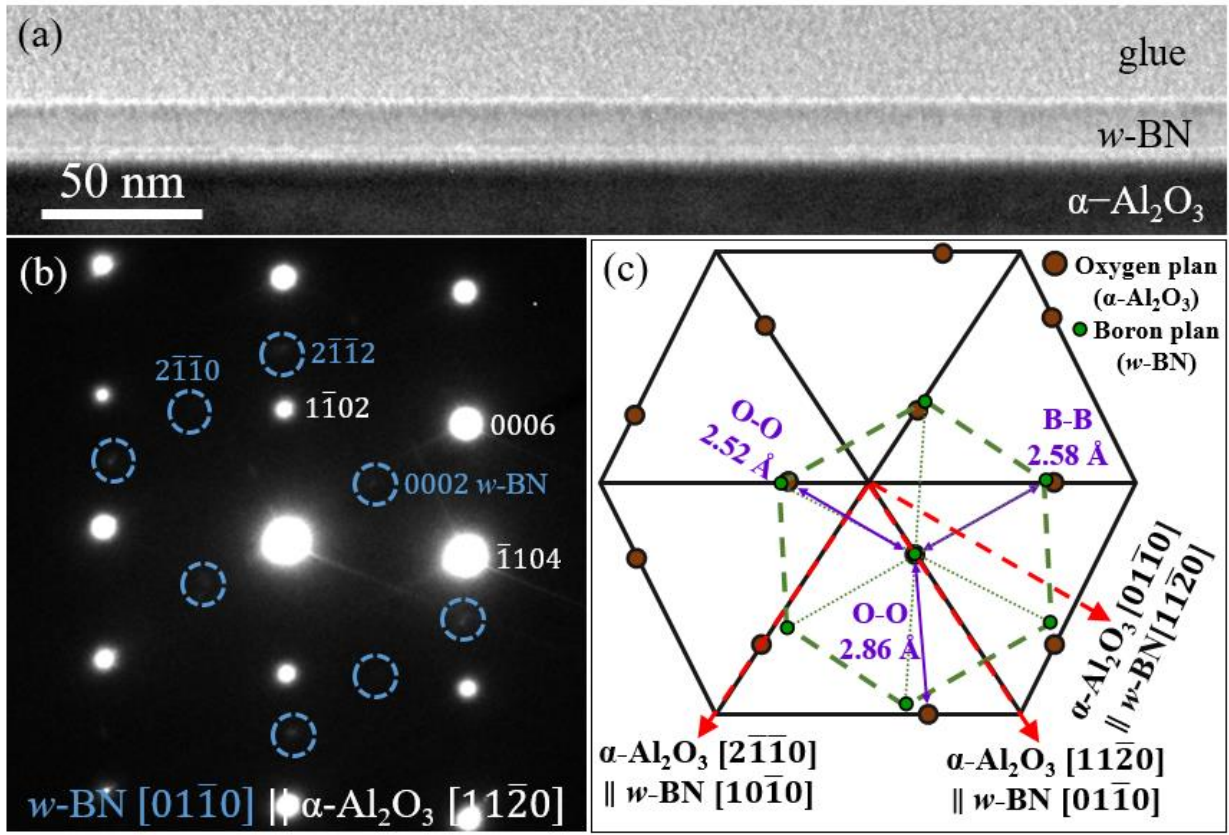

**Figure 2.** (a) TEM bright field image of *w*-BN thin film on *c*-plane sapphire substrate along $<01\bar{1}0>$ Z.A. of *w*-BN. (b) Electron diffraction pattern showing epitaxial relationship between sapphire and *w*-BN is α-Al2O3 $<11\bar{2}0>\|$*w*-BN $<01\bar{1}0>$. (c) Schematic showing in-plane epitaxial relationship between *w*-BN and sapphire substrate.

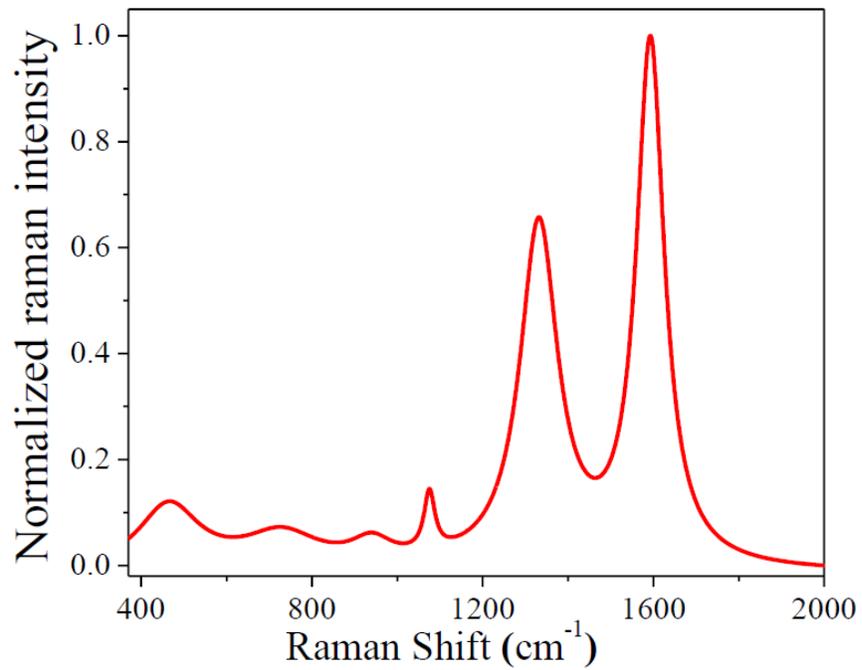

**Figure 3.** Raman spectra from *w*-BN epitaxial thin film.

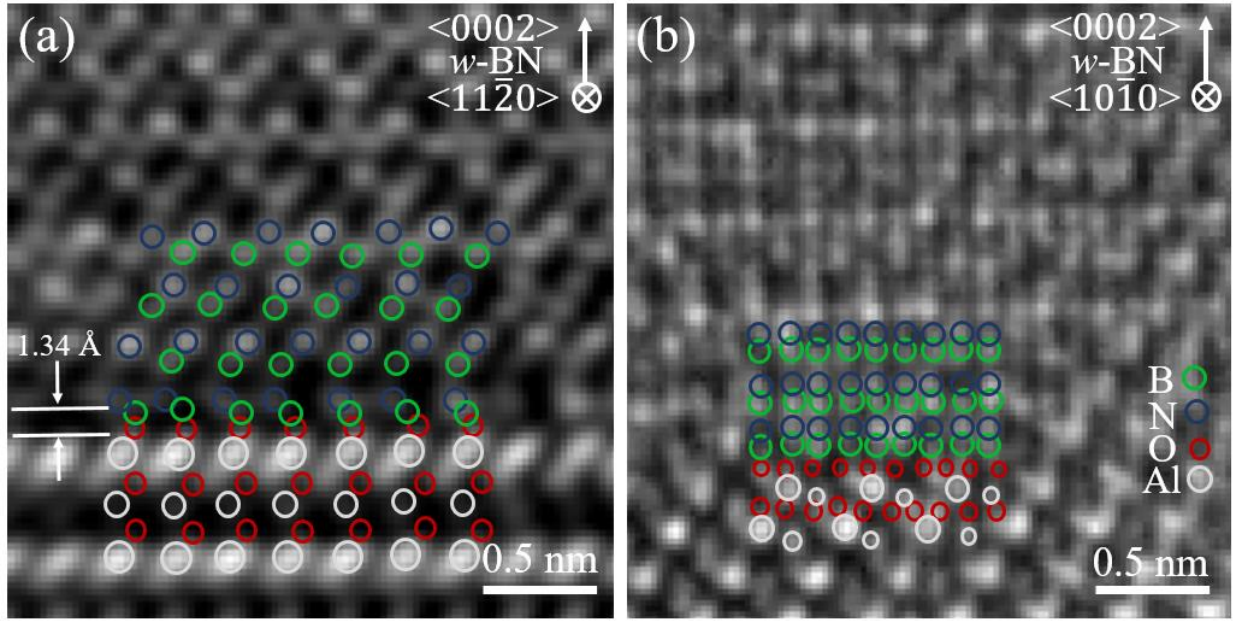

**Figure 4.** (a) & (b) HRTEM image of *w*-BN on c-plane sapphire along two different zone axes of sapphire showing atomic arrangement and registry of atoms at the film substrate interface. The inter-atomic distances are mentioned in the figure.

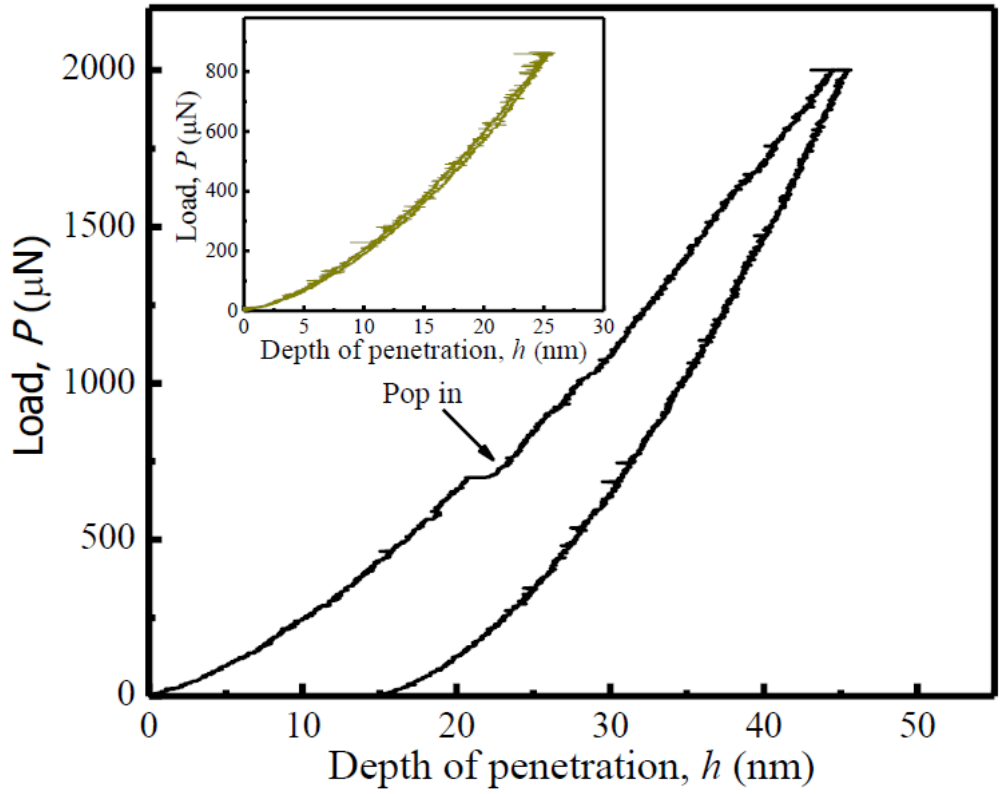

**Figure 5.** Load displacement curves exhibiting pop in due to cracking or delamination phenomenon. (Inset graph shows shallow depth indent without pop-in).

# Chemically stabilized epitaxial wurtzite-BN thin film

Badri Vishal,[1] Rajendra Singh,[1] Abhishek Chaturvedi[2], Ankit Sharma,[1] M.B. Sreedhara,[1] Rajib Sahu,[1] Usha Bhat,[1] Upadrasta Ramamurty[2], and Ranjan Datta[1,*]

[1]*International Centre for Materials Science, Chemistry and Physics of Materials Unit, Jawaharlal Nehru Centre for Advanced Scientific Research, Bangalore 560064, India.*

[2]*Department of Materials Engineering, Indian Institute of Science, Bangalore 560012, India.*

1. **Charge density plot mono and bi-layer *h*-BN**

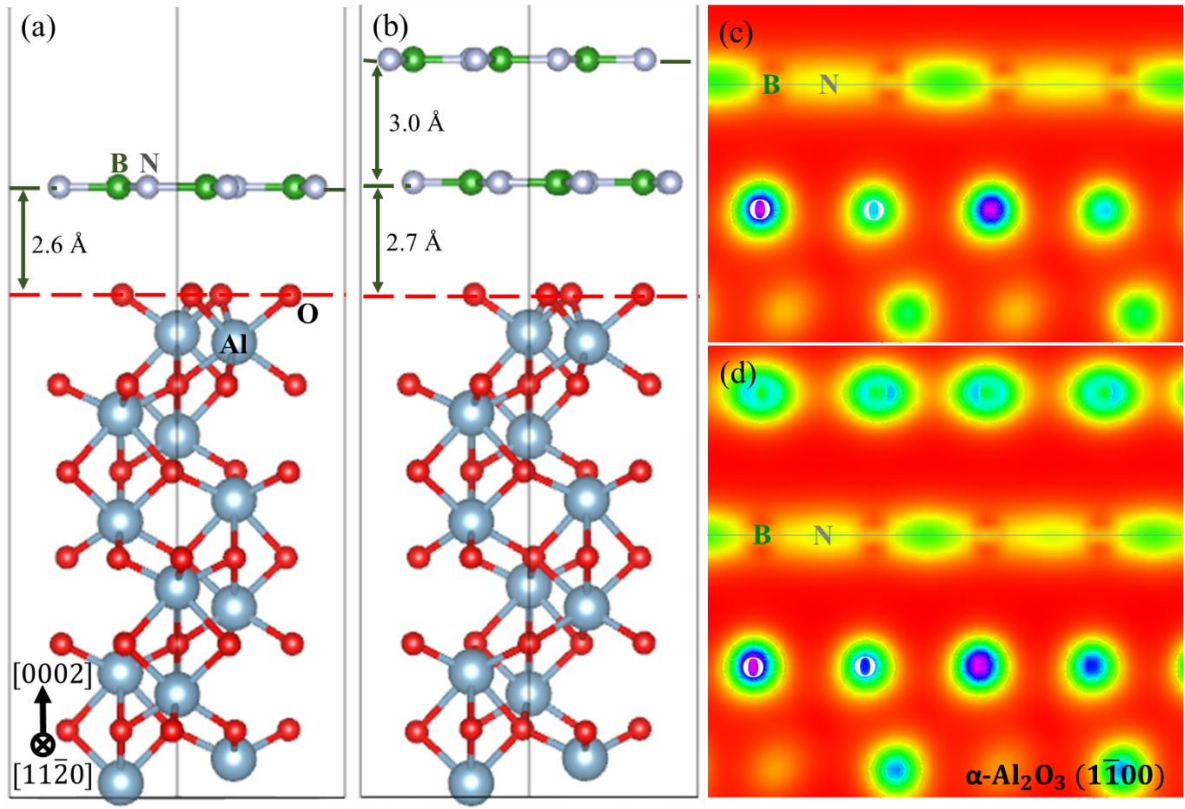

**Figure S1.** Schematic structure of (a) monolayer and (b) bi-layer *h*-BN on c-plane sapphire. Charge density plot showing no additional interaction between B and O atoms at the interface between flat *h*-BN and O-terminating sapphire substrate.

2. **Reference Raman spectra from literature:**

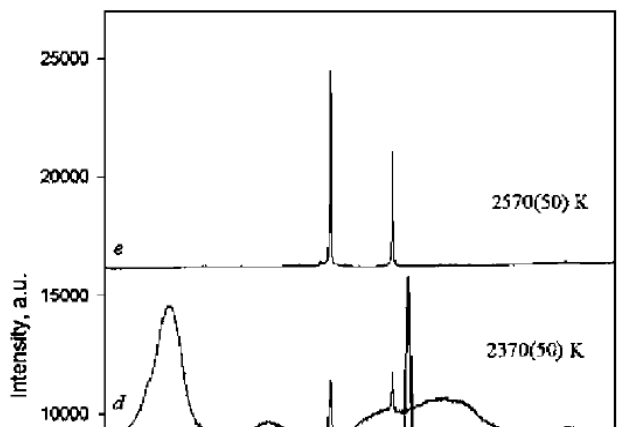

**Figure S2**. Experimental reference Raman spectra for different crystal form of BN.[23] *h*-BN has a unique peak at 1370 cm$^{-1}$, *c*-BN has two sharp peaks at 1057 and 1309 cm$^{-1}$ and *w*-BN has very broad peak around 1600 cm$^{-1}$ along with some associated broad peaks.

3. **X-ray diffraction pattern and XPS results:**

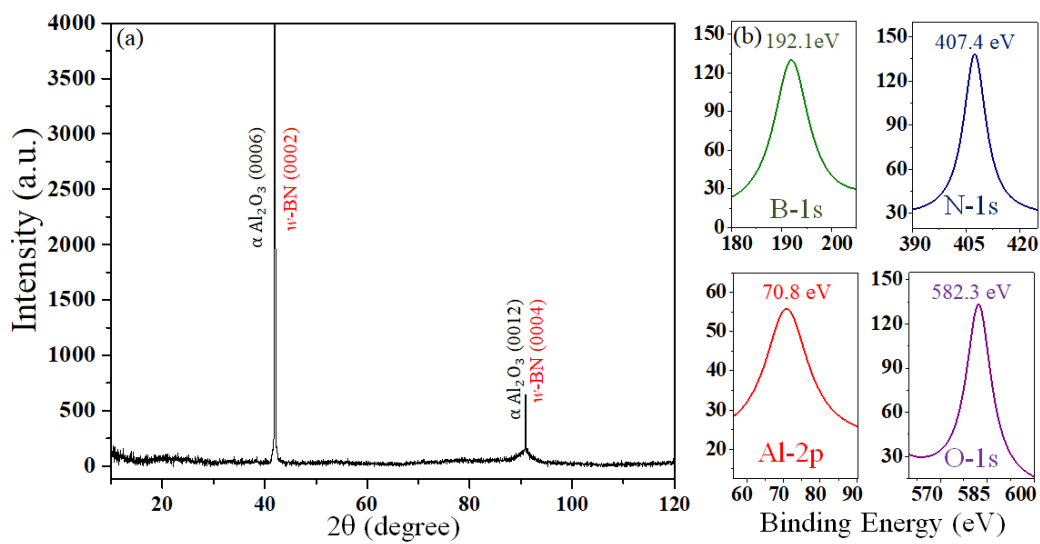

**Figure S3.** (a) X-ray diffraction pattern of epitaxial *w*-BN on *c*-plane sapphire. Kindly note that the (0002) peak of *w*-BN (2θ = 41.94°) superimposes with (0006) of sapphire (2θ = 41.83°). No other peaks corresponding to to *h*-BN, *c*-BN and polycrystalline grains are observed. (b) XPS spectra of B 1s, N 1s, Al 2p and O 1s identifying the elements present both in the film and substrate.

4. **HRTEM**

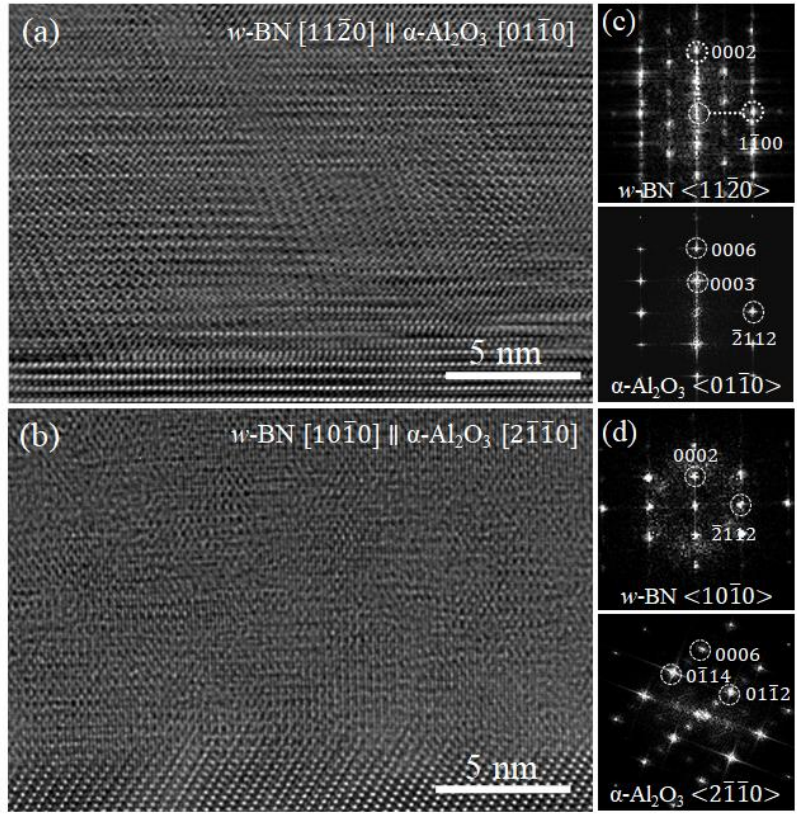

**Figure S4.** HRTEM image of *w*-BN on *c*-plane sapphire along (a) <01$\bar{1}$0>, (b) <2$\bar{1}\bar{1}$0> of sapphire. Corresponding FFT pattern both from the film and substrate areas are given in the (c) and (d), respectively.

## 5. Nano-Indentation

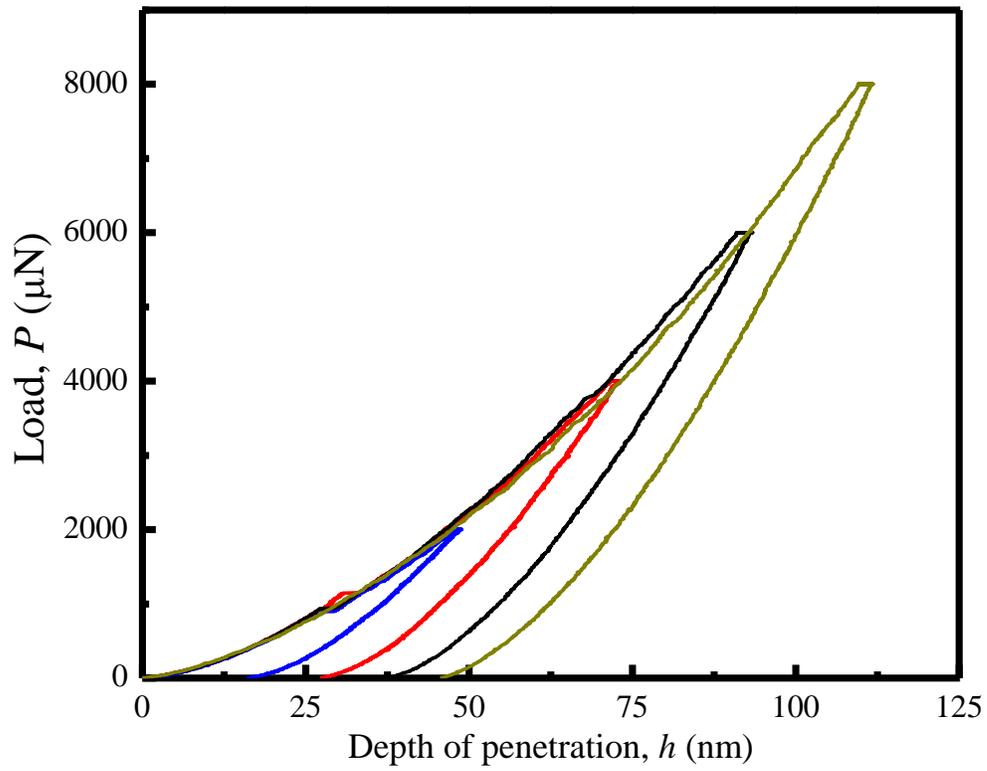

**Figure S5.** Representative indents for α-Al$_2$O$_3$ for various penetration depths.